# The Isotopic Foldy-Wouthuysen Representation as a Possible Key to Understanding the Dark Matter

V.P. Neznamov

FSUE RFNC-VNIIEF, 607188, Sarov, N.Novgorod region

e-mail: Neznamov@vniief.ru

Abstract

The paper considers, within the Standard Model, physical pictures of combinations and interactions of elementary particles, which are allowed within the isotopic Foldy-Wouthuysen representation. A hypothesis is suggested and proved that the properties of the world of dark matter do not contradict the two physical pictures considered in the paper.

To construct the Standard Model without Higgs bosons in the fermion sector, the paper [1] introduces the isotopic Foldy-Wouthuysen representation. Let's describe, in brief, the idea and purpose of this representation.

The system of units with $\hbar = c = 1$ is used below; $x, p, B$ are 4-vectors; $p^\mu = i\dfrac{\partial}{\partial x_\mu}$; the inner product is taken in the form

$$xy = x^\mu y_\mu = x^0 y^0 - x^k y^k, \quad \mu = 0,1,2,3; \quad k = 1,2,3;$$

$$\alpha^\mu = \begin{cases} 1, \mu = 0 \\ \alpha^k, \mu = k = 1,2,3 \end{cases}; \quad \gamma^\mu = \gamma^0 \alpha^\mu; \quad \beta = \gamma^0, \quad \alpha^k, \gamma^\mu, \gamma^5 - \text{Dirac matrices}$$

Consider the density of Hamiltonian of a Dirac particle with mass $m_f$ interacting with an arbitrary abelian boson field $B^\mu$

$$\mathcal{H}_D = \psi^\dagger \left( \vec{\alpha}\vec{p} + \beta m_f + q\alpha_\mu B^\mu \right) \psi = \psi^\dagger (P_L + P_R)\left( \vec{\alpha}\vec{p} + \beta m_f + q\alpha_\mu B^\mu \right)(P_L + P_R)\psi =$$

$$= \psi_L^\dagger \left( \vec{\alpha}\vec{p} + q\alpha_\mu B^\mu \right)\psi_L + \psi_R^\dagger \left( \vec{\alpha}\vec{p} + q\alpha_\mu B^\mu \right)\psi_R + \psi_L^\dagger \beta m_f \psi_R + \psi_R^\dagger \beta m_f \psi_L \qquad (1)$$

In (1) $q -$ is the interaction constant; $P_L = \dfrac{1-\gamma_5}{2}, P_R = \dfrac{1+\gamma_5}{2} -$ are the left and right projective operators; $\psi_L = P_L\psi, \psi_R = P_R \cdot \psi -$ are the left and right components of the Dirac field operator $\psi$.

The reason why the abelian case is considered for the field $B^\mu$ is simplicity. As will be shown below, using a general case of a Dirac particle interacting with non-abelian boson fields would not change the conclusions and implications of this paper. The density of Hamiltonian $\mathcal{H}_D$ can be used to obtain the motion equations for $\psi_L$ and $\psi_R$:

$$\begin{aligned} p_0 \psi_L &= \left( \vec{\alpha}\vec{p} + q\alpha_\mu B^\mu \right)\psi_L + \beta m_f \psi_R \\ p_0 \psi_R &= \left( \vec{\alpha}\vec{p} + q\alpha_\mu B^\mu \right)\psi_R + \beta m_f \psi_L \end{aligned} \qquad (2)$$

Introduce the eight-component field operator $\Phi_1 = \begin{pmatrix} \psi_R \\ \psi_L \end{pmatrix}$ and isotopic matrices $\tau_3 = \begin{pmatrix} I & 0 \\ 0 & -I \end{pmatrix}$, $\tau_1 = \begin{pmatrix} 0 & I \\ I & 0 \end{pmatrix}$ which act on the four upper and four lower components of operator $\Phi_1$.

Hence, equations (2) can be written as

$$p_0 \Phi_1 = \left( \vec{\alpha} \vec{p} + \tau_1 \beta m_f + q\alpha_\mu B^\mu \right) \Phi_1. \tag{3}$$

As far as $\tau_1$ commutes with the right-hand side of equation (3), the field $\Phi_2 = \tau_1 \Phi_1 = \begin{pmatrix} \psi_L \\ \psi_R \end{pmatrix}$ is also a solution to equation (3).

$$p_0 \Phi_2 = \left( \vec{\alpha} \vec{p} + \tau_1 \beta m_f + q\alpha_\mu B^\mu \right) \Phi_2 \tag{4}$$

Finally, use $\Phi_2 = \tau_1 \Phi_1$ and write equations (3), (4) in the following form

$$p_0 \Phi_1 = \left( \vec{\alpha} \vec{p} + \tau_1 \beta m_f + \frac{1}{2} q\alpha_\mu B^\mu \right) \Phi_1 + \frac{1}{2} q\tau_1 \alpha_\mu B^\mu \Phi_2$$

$$p_0 \Phi_2 = \left( \vec{\alpha} \vec{p} + \tau_1 \beta m_f + \frac{1}{2} q\alpha_\mu B^\mu \right) \Phi_2 + \frac{1}{2} q\tau_1 \alpha_\mu B^\mu \Phi_1 \tag{5}$$

One can see from the paper [1] and the description below that being actually equivalent in the Dirac representation, equations (3), (4), and (5) lead to different physical pictures of combinations and interactions of elementary particles in the isotopic Foldy-Wouthuysen representation.

We know from [2] that the Foldy-Wouthuysen transformation is performed with the unitary operator $U_{FW}$. In this case, the Dirac field operator and the Dirac equation's Hamiltonian are transformed as follows:

$$\psi_{FW} = U_{FW} \psi_D$$

$$H_{FW} = U_{FW} H_D U_{FW}^\dagger - i U_{FW} \frac{\partial U_{FW}^\dagger}{\partial t}. \tag{6}$$

Hamiltonian $H_{FW}$ in the Foldy-Wouthuysen representation is block-diagonal relative to the upper and lower components of the field operator $\psi_{FW}$.

The second obligatory condition for transformation to the *FW* representation for free motion and motion in static external fields is the requirement that either the upper components, or lower components of $\psi_{FW}$ become zero [3].

The papers [4], [5] suggest a direct method to construct the Foldy-Wouthuysen transformation in case of fermions interacting with arbitrary boson fields. The transformation matrix $U_{FW}$ and the relativistic Hamiltonian $H_{FW}$ (6) have been obtained as powers series of the coupling constant:

$$U_{FW} = U_{FW}^0 \left(1 + q\delta_1 + q^2\delta_2 + q^3\delta_3 + ....\right)$$

$$H_{FW} = \beta E + qK_1 + q^2 K_2 + q^3 K_3 + .... \qquad (7)$$

In expressions (7), $q$ is the coupling constant, $U_{FW}^0$ is the *FW* representation matrix for free Dirac particles, $E = \sqrt{\vec{p}^2 + m_f^2}$.

Using Hamiltonian (7) in the *FW* representation, quantum electrodynamics and the Standard Model was analyzed, a number of quantum-field effects were calculated, and *SU-2* – invariant derivation of the Standard Model was suggested with initially massive fermions and no Yukawa interactions between Higgs bosons and fermions [4], [6].

The isotopic Foldy-Wouthuysen representation introduced in the paper [1] also allows constructing an *SU-2* – invariant Standard Model with originally massive fermions. In contrast to the standard *FW* transformation (see [2], [4], [5]), the isotopic *FW* transformation does not transform Dirac bi-spinors $\psi_R, \psi_L$ (2), but only performs diagonalization of Hamiltonians relative to the four upper and four lower components of the eight-component fields $\Phi_1, \Phi_2$.

Upon application of one and the same isotopic Foldy-Wouthuysen transformation to equations (3), (4), and (5) we obtain [1]

$$p_0 \Phi_{1FW} = \left( \tau_3 E + qK_1 + q^2 K_2 + q^3 K_3 + ... \right) \Phi_{1FW}$$

$$\mathcal{H}_{FW}^{I} = \Phi_{1FW}^{\dagger} \left( \tau_3 E + qK_1 + q^2 K_2 + q^3 K_3 + ... \right) \Phi_{1FW} \tag{8}$$

$$p_0 \Phi_{2FW} = \left( \tau_3 E + qK_1 + q^2 K_2 + q^3 K_3 + ... \right) \Phi_{2FW}$$

$$\mathcal{H}_{FW}^{II} = \Phi_{2FW}^{\dagger} \left( \tau_3 E + qK_1 + q^2 K_2 + q^3 K_3 + ... \right) \Phi_{2FW} \tag{9}$$

$$p_0 \Phi_{1FW} = \left( \tau_3 E + \frac{q}{2} K_1 + \left(\frac{q}{2}\right)^2 K_2 + \left(\frac{q}{2}\right)^3 K_3 + .... \right) \Phi_{1FW} +$$

$$+ \left( \frac{q}{2} K_{1\tau_1} + \left(\frac{q}{2}\right)^2 K_{2\tau_1} + \left(\frac{q}{2}\right)^3 K_{3\tau_1} + ... \right) \Phi_{2FW}$$

$$p_0 \Phi_{2FW} = \left( \tau_3 E + \frac{q}{2} K_1 + \left(\frac{q}{2}\right)^2 K_2 + \left(\frac{q}{2}\right)^3 K_3 + .... \right) \Phi_{2FW} +$$

$$+ \left( \frac{q}{2} K_{1\tau_1} + \left(\frac{q}{2}\right)^2 K_{2\tau_1} + \left(\frac{q}{2}\right)^3 K_{3\tau_1} + ... \right) \Phi_{1FW} \tag{10}$$

$$\mathcal{H}_{FW}^{IV} = \Phi_{1FW}^{\dagger} \left( \tau_3 E + \frac{q}{2} K_1 + \left(\frac{q}{2}\right)^2 K_2 + \left(\frac{q}{2}\right)^3 K_3 + .... \right) \Phi_{1FW} +$$

$$+ \Phi_{1FW}^{\dagger} \left( \frac{q}{2} K_{1\tau_1} + \left(\frac{q}{2}\right)^2 K_{2\tau_1} + \left(\frac{q}{2}\right)^3 K_{3\tau_1} + .... \right) \Phi_{2FW} +$$

$$+ \Phi_{2FW}^{\dagger} \left( \tau_3 E + \frac{q}{2} K_1 + \left(\frac{q}{2}\right)^2 K_2 + \left(\frac{q}{2}\right)^3 K_3 + .... \right) \Phi_{2FW} +$$

$$+ \Phi_{2FW}^{\dagger} \left( \frac{q}{2} K_{1\tau_1} + \left(\frac{q}{2}\right)^2 K_{2\tau_1} + \left(\frac{q}{2}\right)^3 K_{3\tau_1} + ... \right) \Phi_{1FW}$$

Expressions (8)-(10) also include densities of Hamiltonians corresponding to the resultant expressions in the isotopic Foldy-Wouthuysen representation.

Apparently, it is possible to construct the density of Hamiltonian $\mathcal{H}_{FW}^{III}$ with equations for fields $\Phi_{1FW}$, $\Phi_{2FW}$ taken from expressions (8), (9):

$$\mathcal{H}_{FW}^{III} = \mathcal{H}_{FW}^{I} + \mathcal{H}_{FW}^{II}. \tag{11}$$

Basic orthonormal functions $\Phi_{1FW}(x)$, $\Phi_{2FW}(x)$ for free motion of fermions can be expressed in terms of the left and right components of the Dirac field $\psi(x)$ and have the form [1]

$$\begin{aligned}
\Phi_{1FW}^{(+)}(\vec{x},t) &= U_{1FW}^{0}\Phi_{1}^{(+)}(\vec{x},t) = e^{-iEt}\begin{pmatrix} \sqrt{\dfrac{2E}{E+\vec{\sigma}\vec{p}}}\psi_{R}^{(+)}(\vec{x}) \\ 0 \end{pmatrix} \\
\Phi_{1FW}^{(-)}(\vec{x},t) &= U_{FW}^{0}\Phi_{1}^{(-)}(\vec{x},t) = e^{iEt}\begin{pmatrix} 0 \\ \sqrt{\dfrac{2E}{E+\vec{\sigma}\vec{p}}}\psi_{L}^{(-)}(\vec{x}) \end{pmatrix} \\
\Phi_{2FW}^{(+)}(\vec{x},t) &= U_{FW}^{0}\Phi_{2}^{(+)}(\vec{x},t) = e^{-iEt}\begin{pmatrix} \sqrt{\dfrac{2E}{E-\vec{\sigma}\vec{p}}}\psi_{L}^{(+)}(\vec{x}) \\ 0 \end{pmatrix} \\
\Phi_{2FW}^{(-)}(\vec{x},t) &= U_{FW}^{0}\Phi_{2}^{(-)}(\vec{x},t) = e^{iEt}\begin{pmatrix} 0 \\ \sqrt{\dfrac{2E}{E-\vec{\sigma}\vec{p}}}\psi_{R}^{(-)}(\vec{x}) \end{pmatrix}
\end{aligned} \tag{12}$$

In the presence of static external boson fields, basis functions in the Foldy-Wouthuysen representation are similar in their isotopic structure to those in expressions (12). When solving applied problems in the quantum field theory using the theory of perturbations, fermion fields are expanded in a series over solutions to Dirac equations for free motion or motion in static external fields. In our case of the isotopic Foldy-Wouthuysen representation we can expand fermion fields over the basis (12), or over the basis of solutions to Foldy-Wouthuysen equations in static external fields.

Hamiltonians in (8)-(10) are, by definition, diagonal with respect to the upper and lower isotopic components and based on the above considerations they are *SU-2* –invariant, regardless of whether the fermions are massive or massless.

Now, consider the Hamiltonian densities and equations (8)-(10) from the viewpoint of combinations and interactions of elementary particles. Remember that equations (8)-(10) have been obtained from one and the same Dirac equation written in various forms in expressions (3)-(5). Hence, the physical pictures corresponding to equations (8)-(10) can take place in our universe.

First, consider Hamiltonian density $\mathcal{H}_{FW}^{IV}$ and equations (10). The appropriate physical picture is symbolically represented in Fig. 1.

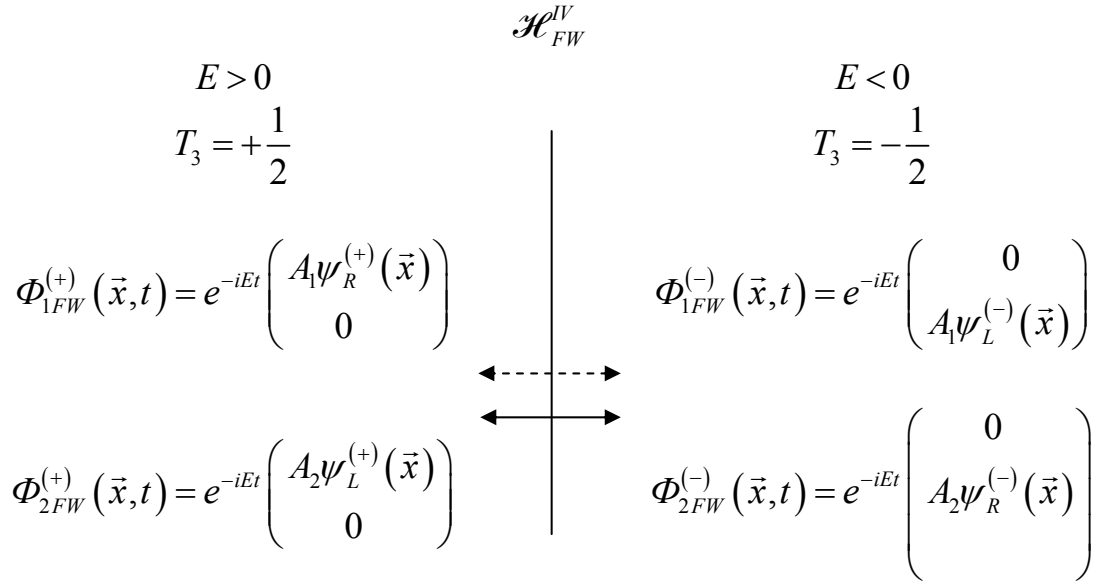

Fig. 1

The left half-plane in Fig. 1 represents states of basic functions (12) with isotopic spin $T_3 = +\frac{1}{2}$ and positive value of energy $E > 0$; the right half-plane represents states with $T_3 = -\frac{1}{2}$ and negative value of energy $E < 0$. Hamiltonian $H^{IV}$ contains the states of the left and right fermions, as well as the states of the left and right antifermions; particles and antiparticles interact with each other both really (solid line with arrows in Fig. 1) and virtually (dashed line with arrows in Fig. 1). The physical pattern in Fig. 1 represents the real world around us.

Now, consider Hamiltonian density $\mathcal{H}_{FW}^{IV}$ with equations from (8), (9). The symbolic representation of combinations and interactions of elementary particles is given in Fig. 2.

$$\mathcal{H}_{FW}^{III}$$

$E > 0$ $\qquad\qquad\qquad\qquad$ $E < 0$

$T_3 = +\dfrac{1}{2}$ $\qquad\qquad\qquad\qquad$ $T_3 = -\dfrac{1}{2}$

$$\Phi_{1FW}^{(+)}(\vec{x},t) = e^{-iEt}\begin{pmatrix} A_1\psi_R^{(+)}(\vec{x}) \\ 0 \end{pmatrix} \qquad \Phi_{1FW}^{(-)}(\vec{x},t) = e^{-iEt}\begin{pmatrix} 0 \\ A_1\psi_L^{(-)}(\vec{x}) \end{pmatrix}$$

$$\Phi_{2FW}^{(+)}(\vec{x},t) = e^{-iEt}\begin{pmatrix} A_2\psi_L^{(+)}(\vec{x}) \\ 0 \end{pmatrix} \qquad \Phi_{2FW}^{(-)}(\vec{x},t) = e^{-iEt}\begin{pmatrix} 0 \\ A_2\psi_R^{(-)}(\vec{x}) \end{pmatrix}$$

Fig. 2

The world in Fig. 2 is poorer than that in Fig.1. The physical picture in Fig. 2 has both left and right fermions and left and right antifermions. However, there are no interactions between real fermions and antifermions; there is only virtual interaction between them (a dashed line with arrows in Fig. 2). Fig. 2 allows strong and electromagnetic interactions between particles and antiparticles without any real interactions between them. There are no processes of production and absorption of real particle-antiparticle pairs and there are no coupled states of real particles and antiparticles, etc. Weak interactions are significantly poorer, as well, because there are no processes of simultaneous production and absorption of real particles together with antiparticles.

Note that vacuum states are identical in Fig1 and Fig. 2.

Consider Hamiltonian densities $\mathcal{H}_{FW}^{I}$, $\mathcal{H}_{FW}^{II}$ with the corresponding equations from (8), (9). The picture is symbolically shown in Figs. 3, 4.

$$\mathcal{H}_{FW}^{I}$$

| $E > 0$ | $E < 0$ |
|---|---|
| $T_3 = +\dfrac{1}{2}$ | $T_3 = -\dfrac{1}{2}$ |
| $\Phi_{1FW}^{(+)}(\vec{x},t) = e^{-iEt}\begin{pmatrix} A_1\psi_R^{(+)}(\vec{x}) \\ 0 \end{pmatrix}$ | $\Phi_{1FW}^{(-)}(\vec{x},t) = e^{-iEt}\begin{pmatrix} 0 \\ A_1\psi_L^{(-)}(\vec{x}) \end{pmatrix}$ |

Fig. 3

$$\mathcal{H}_{FW}^{II}$$

| $E > 0$ | $E < 0$ |
|---|---|
| $T_3 = +\dfrac{1}{2}$ | $T_3 = -\dfrac{1}{2}$ |
| $\Phi_{2FW}^{(+)}(\vec{x},t) = e^{-iEt}\begin{pmatrix} A_2\psi_L^{(+)}(\vec{x}) \\ 0 \end{pmatrix}$ | $\Phi_{2FW}^{(-)}(\vec{x},t) = e^{-iEt}\begin{pmatrix} 0 \\ A_2\psi_R^{(-)}(\vec{x}) \end{pmatrix}$ |

Fig. 4

It follows from Fig. 3 and Fig. 4 that Hamiltonian densities $\mathcal{H}_{FW}^{I}$, $\mathcal{H}_{FW}^{II}$ stipulate existence of either right fermions and left antifermions ($\mathcal{H}_{FW}^{I}$), or left fermions and right antifermions ($\mathcal{H}_{FW}^{II}$). In both cases, no interactions occur between real particles and antiparticles.

The physical world in Fig. 3 or in Fig. 4 has no electromagnetic and no strong interactions, since because of parity preservation in such interactions, both left and right fermions are required to be present. The same requirement is valid for processes with neutral current weak interactions and so, they are also absent in Figs. 3 and 4. The processes with weak charqed currents and with participation of either left fermions (Fig. 4), or left antifermions (Fig. 3) also appear to be suppressed because of impossibility to emit, or absorb particles and antiparticles together. The vacuum state in Fig. 3 and Fig. 4 significantly differs from that in Fig. 1 and Fig. 2. Since no interactions are allowed (except for the gravitational interaction) in the vacuums in Fig. 3 and Fig. 4, no "bouillon" of particle-antiparticle virtual pairs and virtual carriers of interactions is found.

Here is a summary of the physical pictures given in Figs. 1, 2, 3, and 4.

As the strong and electromagnetic interactions are prohibited and weak interactions are virtually absent, the world shown in Figs. 3, 4 has the following properties:

- it does not emit/absorb light;
- it is electrically neutral;
- the motion is non-relativistic;
- it weakly interacts with the outer world.

The aforementioned properties are those of dark matter that was discovered during the last quarter of the past century (see, for example, [7]) and constitutes ~ 26% of the universe. Hence, we may assume that dark matter is an implementation of the physical pictures shown in Fig.3 and Fig.4. It consists of either right fermions and left antifermions, or left fermions and right antifermions. A set of fermions and antifermions requires no changes in the composition of particles in the Standard Model.

As it has been mentioned earlier, the physical picture in Fig.1 represents the world around us in our part of the universe. The baryon matter ("light matter") constitutes ~ 4% of the universe structure. If we assume that in the past, transformation from our picture (see Fig.1) to the physical picture shown in Figs.3 and 4 occurred in a part of the universe, then, in addition to dark matter generation, vacuum would have been significantly restructured owing to the disappearance of quark, gluon, and electroweak condensates. This leads us to a speculation that such restructuring is associated with the problem of dark energy, which currently constitutes ~70% of the universe according to observation data [7].

A principal question still requires an answer:

- How and why does the universe, if it does at all, transform to various physical pictures of its elementary particle combinations and interactions?

It should be noted once again that different physical pictures of the composition and interactions of elementary particles have been derived basing on one equation of the Dirac field interacting with boson fields described by equations (3)-(5) using the isotopic Foldy-Wouthuysen transformation that allows writing equations of fields with massive fermions and their Hamiltonians in the form invariant relative to $SU$-2 – transformations.

The combinations of elementary particles in the physical pictures described above is within the set of particles of the Standard Model.